\documentclass{article}
\usepackage[utf8]{inputenc}
\usepackage{amsmath,amssymb,pxfonts}
\usepackage{booktabs}
\usepackage{pdflscape}
\usepackage{algpseudocode}
\usepackage{algorithm}
\usepackage{hyperref}
\newcommand{\nothing}[1]{}
\usepackage{graphicx}
\newtheorem{prop}{Proposition}
\newtheorem{rem}{Remark}
\newtheorem{thm}{Theorem}
\newtheorem{cor}{Corollary}

\title{Model-free Hedging of Impermanent Loss \\ in Geometric Mean Market Makers}
\author{Masaaki Fukasawa\thanks{Graduate School of Engineering Science, Osaka University, Japan}, Basile Maire\thanks{Desma Eight, LLC, NY, USA}, and Marcus Wunsch\thanks{ZHAW School of Management and Law, Switzerland}}

\begin{document}

\maketitle

\begin{abstract}
We consider Geometric Mean Market Makers -- a special type of Decentralized Exchange -- with two types of users: liquidity takers and arbitrageurs. 
Liquidity takers trade at prices that can create arbitrage opportunities, while arbitrageurs align the exchange's price with the external market price. 
We show that in Geometric Mean Market Makers charging proportional transaction fees, Impermanent Loss can be super-hedged by a model-free rebalancing strategy. 
Moreover, we demonstrate that in such a DEX, the exchange rate is of finite variation, so that loss-versus-rebalancing (the shortfall of providing liquidity versus the corresponding constant-weights portfolio) vanishes. 
\end{abstract}
\\



\newpage
\section{Introduction}
\subsection{Geometric Mean Market Makers}
In Decentralized Exchanges (DEX) such as Uniswap, a Liquidity Pool (''LP'') consists of a pair of digital assets that are deposited as reserves by liquidity providers. 
The Automated Market Maker (AMM) associated with this pool is a smart contract that executes orders from liquidity takers. 
A general class of AMMs are so-called Constant Function Market Makers (CFMM), where liquidity takers swap $\Delta x$ units of the first currency for $\Delta y$ units of the second currency in accordance with the equation
\begin{equation}\label{CFMM}
   \varphi(x,y, \Delta x,\Delta y) = \varphi(x,y,0,0),
\end{equation}
where $\varphi$, called the \textit{trading function},
is a function which is increasing with respect to each of its arguments,
and $x>0$ and $y>0$ are, respectively, the reserve amounts of the first and second currency in the LP at the time.
When $\Delta x >0$ ($\Delta x < 0$), this means a liquidity taker adds $\Delta x$  units of the first ($\Delta y$ of the second) currency to the LP to receive $-\Delta y > 0$  units of the second ($-\Delta x > 0$ of the first) currency from the LP. 
The reserves in the LP are then updated from $(x,y)$ to $(x + \Delta x, y + \Delta y)$.
We call the infinitesimal exchange ratios
\begin{equation}\label{BA}
    \lim_{\Delta x \uparrow 0} - \frac{\Delta y}{\Delta x}, \ \ \lim_{\Delta x \downarrow 0} - \frac{\Delta y}{\Delta x},
\end{equation}
respectively, the ask and bid prices of the first currency.
These prices depend on the pool's reserves $(x,y)$ at the time.

We focus on Geometric Mean Market Makers (G3M), a particular form of CFMMs
employed by popular AMM-protocols such as Uniswap and Balancer.
G3Ms are defined by the following rule: the reserves of the AMM before and after each trade must have the same weighted geometric mean. 
\\
We consider G3Ms charging proportional fees:
\begin{equation}\label{eq:CD}
 (x + (1-\tau H(\Delta x)) \Delta x)^\alpha(y+ (1-\tau H(\Delta y))\Delta y)^{1-\alpha} = x^\alpha y^{1-\alpha},
\end{equation}
where $\alpha \in (0,1)$, $\tau \in (0,1)$ and $H$ is the Heaviside function.
This means that $\tau$ denotes the fee charged on the incoming currency.
The rule \eqref{eq:CD} corresponds to \eqref{CFMM} with
the trading function
\begin{equation*}
    \varphi(x,y,\Delta x, \Delta y)
    = (x + (1-\tau H(\Delta x)) \Delta x)^\alpha(y+ (1-\tau H(\Delta y))\Delta y)^{1-\alpha}.
\end{equation*}
The ask and bid prices \eqref{BA} are then, respectively,
\begin{equation}\label{eq:BA2}
    \lim_{\Delta x \uparrow 0} - \frac{\Delta y}{\Delta x}
    = \frac{1}{1-\tau} \frac{\alpha}{1-\alpha} \frac{ y}{x}, \ \ \lim_{\Delta x \downarrow 0} - \frac{\Delta y}{\Delta x} = (1-\tau)  \frac{\alpha}{1-\alpha} \frac{ y}{x},
\end{equation}
where $(x,y)$ are the current LP's reserves. 

\subsection{Impermanent Loss}

Throughout this article, we assume interest rates to be zero.
Suppose also that there are external liquid exchanges\footnote{These could be, for instance, Centralized Exchanges such as Binance or Kraken, or other DEX.} for the currency pair, and denote by $S^\ast = \{S^\ast_t\}$ the price (exchange ratio) process. More precisely, $S^\ast_t$ is the 
external price at time $t$ for one unit of the first currency in terms of the second. 
For simplicity, we assume that there is only one hold-to-maturity liquidity provider, 
i.e., no liquidity pool 
withdrawals and additions over time, an assumption that comes without loss 
of generality when $\tau=0$ as seen in \cite{FMW}.

The value of the LP at time $t$ is evaluated in terms of the second currency as
\begin{equation*}
V_t := Y_t  + X_t S^\ast_t,
\end{equation*}
where $(X_t,Y_t)$ are the reserves at time $t$ in the LP.
Due to being short gamma~\cite{FMW}, the liquidity provider faces a risk called Impermanent Loss (IL; synonymously, divergence loss) defined as
\begin{equation*}
    \Psi_t =  Y_0 + X_0 S^\ast_t - V_t,
\end{equation*}
where $Y_0 + X_0 S^\ast_t$ is the value of a buy-and-hold strategy with the same 
initial value $V_0$ (so that $\Psi_0 = 0$).
An interpretation of IL is the mark-to-market loss for the liquidity provider who financed  the initial reserve $X_0 > 0$ of the first currency at time $0$ by borrowing:
\begin{equation*}
    \Psi_t = V_0- V_t + X_0(S^\ast_t-S^\ast_0).
\end{equation*}
The case of no fee ($\tau=0$) has been already studied by several authors, see Section~\ref{sec:review} below. 
As revisited in Remark~\ref{rem1}, we know that under the no-arbitrage
condition and zero fees, we can express the IL as a
strictly convex function $\psi$ of the external market price,
and therefore, by Jensen's inequality, its risk neutral expectation implies a positive expected IL:
\begin{equation}\label{Jensen}
    \mathsf{E}_Q[\Psi_t] = \mathsf{E}_Q[\psi(S^\ast_t)]
    > \psi\left(\mathsf{E}_Q[S^\ast_t]\right)
    = \psi(S^\ast_0) = \Psi_0 = 0,
\end{equation}
where $Q$ is any martingale measure of $S^\ast$.
It has been argued that a fee $\tau >0$ is necessary to compensate IL; however, the structure of IL under fee has so far not been well-understood. 
As noted in~\cite{LT}, in the presence of fees, due to the bid-ask spread, the value of LP depends not only on the external price but also on the whole history of trades in the LP.

In our model, there are two types of traders that interact with the AMM: 
\emph{liquidity takers} and \emph{arbitrageurs}. 
\begin{itemize}
    \item \emph{Liquidity takers} exchange tokens and their trade might result in arbitrage opportunities, that is, the external price $S^\ast_t$ might fall outside the bid-ask spread after the trade.
    \item \emph{Arbitrageurs} seek to optimize their profit by trading between the external market and the AMM so that their profit is maximized. As a result of their activities, the LP's bid-ask range is repositioned so that it encloses the external market price: $B_t \leq S^\ast_t \leq A_t$, where from 
    Equation~\eqref{eq:BA2}:
    \begin{equation*}
    (B_t,A_t) = \left(
(1-\tau)  \frac{\alpha}{1-\alpha} \frac{ Y_t}{X_t}, 
     \frac{1}{1-\tau} \frac{\alpha}{1-\alpha} \frac{ Y_t}{X_t} \right).
    \end{equation*}
\end{itemize}

\subsection{Contribution of this article}

This article can be regarded as an extension of the research presented in~\cite{FMW} in that it relaxes the assumptions on the exchange rate process and extends the analysis to G3Ms with transaction fees. 
Under the assumptions outlined above, we 
develop a stochastic model of which the well-definedness is proved in light of the Skorokhod problem, and 
derive a model-free upper bound for IL.
The upper bound makes no assumptions on the dynamics of the external price process, other than
${S^\ast_t}$ being a positive continuous process, hence the term ''model-free''.
The upper bound corresponds to a rebalancing strategy in ${S^\ast_t}$, and hence,
IL for G3Ms can be super-hedged with the strategy that we describe.

We perform simulation experiments over a fixed period of historical BTC-price data to proxy the external market price $S^\ast_t$. The experiments differ in the fee size, and the trading probability of the different agents. 
The agents are small liquidity takers (trades result in no-arbitrage), large liquidity takers that are immediately followed by optimal arbitrage trades so that the no-arbitrage condition holds, and arbitrage traders that either follow a large liquidity taker or follow an external price move. 
In contrast to the theory, the simulation is based on discretely sampled data, which amounts to dealing with a discontinuous price process.
As a result, our super-hedge is not perfect and only approximative in the simulation results.
Even though the size of $\tau$ is irrelevant to the super-hedge in the theory, not surprisingly, we find that the IL decreases with higher fees in the simulation. 
When large trades dominate, arbitrage trades follow, creating higher fee income and lower IL.

To the best of our knowledge, we present the first results on model-free hedging of Impermanent Loss in Geometric Mean Market Makers - so far, they have been analyzed under the assumption that the underlying dynamics follow geometric Brownian motions (cf. the references in the literature review below). 

\subsection{Literature review}\label{sec:review}
For a general introduction to AMMs, we refer to the survey paper~\cite{M} and the references therein. 
Prior work has mostly focused on theoretical AMMs that do not charge transaction fees. 
For example,~\cite{E} shows that G3Ms without fees underperform equivalent constant-mix portfolios due to arbitrageurs. 
This difference is termed "loss-versus-rebalancing" in~\cite{MMRZ}, who decompose liquidity provision into a component subject to market risk and another component subject to arbitrage trading profits under the assumption that no fees are imposed, while~\cite{CSSS} derive an upper bound for no-arbitrage fee income from liquidity provision using arguments similar to~\cite{ACEL}. 
These articles are complemented by analogous analyses in~\cite{CDM} for concentrated liquidity provision as in \href{https://uniswap.org/blog/uniswap-v3}{Uniswap v3}. 
Finally, in~\cite{AKCNC}, the authors investigate the trade-off that liquidity providers face when including fees: while fees add to the portfolio value of a liquidity provider, they also reduce trading activity. 

Note that there are several trading venues operating as G3Ms, such as \href{https://app.balancer.fi/#/}{Balancer}, \href{https://uniswap.org/blog/uniswap-v2}{Uniswap v2}, or \href{https://www.sushi.com/earn}{Sushiswap}. 

\subsection{Structure of this article}
The paper is organized as follows. 
In Section~\ref{sec:absenceManipulation}, we show that the trader cannot profit from placing two orders in opposite directions (buy/sell), a desirable property of an AMM. 
We derive the optimal arbitrage strategy in Section~\ref{OA}. 
Section~\ref{sec:reserves} investigates the dynamics of the reserves, which allows us to find an upper bound of the IL in Section~\ref{sec:imploss}.
Finally, Section~\ref{sec:sim} contains a simulation analysis.

\section{Preliminary analysis}
\subsection{The absence of price manipulation}\label{sec:absenceManipulation}
For the G3M under consideration, a block trade is more efficient than split trades as shown in~\cite{AKCNC}. 
We now consider trades with opposite directions.
Denote
\begin{equation*}
    \beta = \frac{\alpha}{1-\alpha}
\end{equation*}
for brevity.
If $\Delta x> 0$, or, equivalently, $\Delta y < 0$, then
 \begin{equation*}
     -\Delta y = f_+(x,y,\Delta x):=y \left(1 - \left( \frac{x}{x + (1-\tau)\Delta x}\right)^{\beta} \right)
\approx 
    (1-\tau) \frac{ \beta y}{x} \Delta x
 \end{equation*}
by \eqref{eq:CD}.
 If on the other hand, $\Delta y > 0$ or equivalently $\Delta x < 0$, then
\begin{equation*}
     -\Delta y = f_-(x,y,\Delta x):=\frac{y}{1-\tau} \left(
1 - \left( \frac{x}{x + \Delta x}\right)^{\beta} \right)
\approx 
    \frac{1}{1-\tau}  \frac{\beta y}{x}\Delta x.
 \end{equation*}
Note that $f_\pm(x,y,0) = 
0$. Denote
\begin{equation*}
    f(x,y,z) =
    \begin{cases}
    f_+(x,y,z) & z \geq 0, \\
     f_-(x,y,z) & z < 0.
     \end{cases}
\end{equation*}
 
\begin{prop}\label{prop1}
 If $\Delta x_1 \Delta x_2 < 0$, then 
 \begin{equation*}
- \Delta y_1 - \Delta y_2  <  - \Delta y,
 \end{equation*}
 where 
\begin{equation*}
\begin{split}
& \Delta y_1 = - f(x,y,\Delta x_1), \\ 
&\Delta y_2 = -f(x+\Delta x_1, y+ \Delta y_1, \Delta x_2),
    \\ 
    & \Delta y = - f(x,y,\Delta x)
    \end{split}
\end{equation*}
and $\Delta x = \Delta x_1 +\Delta x_2$.
\end{prop}
{\it Proof: } See Appendix~\ref{appA}.\\

Proposition~\ref{prop1} shows that the trader cannot
profit from placing two orders in opposite directions.
Some explicit computations are possible when $\alpha = 1/2$ (or equivalently, $\beta = 1$).
For example,  when a liquidity taker wants to sell the amount $\Delta x > 0$ of the first currency,
 the block order gives the amount
\begin{equation*}
-\Delta y = (1-\tau) \frac{y \Delta x}{x+ (1-\tau)\Delta x} > 0
\end{equation*}
of the second currency, 
while by decomposing it as $\Delta x = \Delta x_1 + \Delta x_2$ with $\Delta x_1 > \Delta x$ and $\Delta x_2 < 0$, she gets
\begin{equation*}
   -\Delta y_1 = (1-\tau) \frac{y \Delta x_1}{x+ (1-\tau)\Delta x_1} > 0
\end{equation*}
plus
\begin{equation*}
    -\Delta y_2 = \frac{1}{1-\tau} \frac{( y + \Delta y_1) \Delta x_2}{x+ \Delta x_1 + \Delta x_2} < 0.
\end{equation*}
Observe that
\begin{equation*}
    -\Delta y_1    -\Delta y_2 - (-\Delta y)
    = \frac{\tau x y\Delta x_2 (x + (1-\tau)(x+\Delta x))}{(1-\tau)(x + \Delta x)(x + (1-\tau)\Delta x_1)(x + (1-\tau)\Delta x)} <  0.
\end{equation*}

\subsection{The optimal arbitrage}\label{OA}
Here we describe the optimal arbitrage trade for arbitrageurs when the no arbitrage bound is violated, and show that the no arbitrage bound is recovered as a result of the trade.
Denote by
\begin{equation*}
    a = \frac{1}{1-\tau} \frac{\beta y}{x}, \ \ 
    b = (1-\tau) \frac{\beta y}{x}
\end{equation*}
the ask and bid prices (see \eqref{eq:BA2}).
 Denoting by $s^\ast$  the external price, an arbitrageur's profit is
 \begin{equation*}
     -\Delta y - s^\ast \Delta x
 \end{equation*}
 for an arbitrage order $(\Delta x,\Delta y)$.
 If $s^\ast < b$,
 then $\Delta x > 0$ and the optimal order is the solution of
 \begin{equation*}
     s^\ast = \frac{\partial f_+}{\partial z}(x,y,z)|_{z = \Delta x} =
     b\left(1  + (1-\tau)\frac{\Delta x}{x}\right)^{-\beta-1}
     =(1-\tau)\beta \frac{y + \Delta y}{x + (1-\tau)\Delta x},
 \end{equation*}
 that is, in terms of $\Delta x>0$,
 \begin{equation*}
     \Delta x = \frac{x}{1-\tau}\left( \left(\frac{b}{s^\ast}\right)^{1-\alpha} -1\right).
 \end{equation*}
 For such $(\Delta x,\Delta y)$, we have
 \begin{equation*}
 \beta \frac{y + \Delta y}{x + \Delta x} > s^\ast > (1-\tau) \beta \frac{y + \Delta y}{x + \Delta x}.
 \end{equation*}
If $s^\ast > a$,
 then $\Delta x < 0$ and the optimal order is the solution of
 \begin{equation*}
     s^\ast = \frac{\partial f_-}{\partial z}(x,y,z)|_{z = \Delta x} 
     = a \left(1+\frac{\Delta x}{x}\right)^{-\beta-1}
     = \frac{\beta}{1-\tau} \frac{y + (1-\tau)\Delta y}{x + \Delta x},
 \end{equation*}
 that is, in terms of $\Delta y>0$,
 \begin{equation*}
     \Delta y = \frac{y}{1-\tau}\left( \left(\frac{s^\ast}{a}\right)^{\alpha} -1\right).
 \end{equation*}
 For such $(\Delta x,\Delta y)$, we have
 \begin{equation*}
 \frac{\beta}{1-\tau} \frac{y + \Delta y}{x + \Delta x} > s^\ast > \beta \frac{y + \Delta y}{x + \Delta x}.
 \end{equation*}
 Consequently, if $s^\ast \notin [b,a]$ and an arbitrageur places the optimal order
 $(\Delta x, \Delta y)$ to the LP, the ask and bid prices of the LP are updated so that the no-arbitrage condition is recovered.

\section{The dynamics of the reserves}\label{sec:reserves}
Denote by $(X_t,Y_t)$ the reserves in a LP at time $t$.
A liquidity provider puts initial reserves $(X_0,Y_0)$ at time $0$.
We regard $X = \{X_t\}$ and $Y = \{Y_t\}$ as right-continuous stochastic processes with left-limits.
Their left continuous modifications $X_- = \{X_{t-}\}$ and $Y_- = \{Y_{t-}\}$
are defined as $X_{t-} = \lim_{s\uparrow t}X_s$
and
$Y_{t-} = \lim_{s\uparrow t}Y_s$.
By definition, $\Delta X = X - X_-$ and $\Delta Y = Y-Y_-$.
The ask and bid price processes $A = \{A_t\}$ and $B=\{B_t\}$ are respectively defined as
\begin{equation*}
\begin{split}
   &A =  \lim_{\Delta x \uparrow 0}  \frac{f_-(X,Y,\Delta x)}{\Delta x} = \frac{1}{1-\tau}S, 
   \\& B = \lim_{\Delta x \downarrow 0} 
   \frac{f_+(X,Y,\Delta x)}{\Delta x} = (1-\tau)S,
   \end{split}
\end{equation*}
where $S = \{S_t\}$ is defined by
\begin{equation*}
    S = \frac{\beta Y}{X}.
\end{equation*}
We assume the external price process $S^\ast = \{S^\ast_t\}$ to be continuous. To avoid an immediate arbitrage trade, the initial reserves should satisfy
$B_0 \leq S^\ast_0 \leq A^\ast_0$.\\

There are three types of transactions that drive the reserves process $(X,Y)$.
The first type is by a \emph{small order} from a liquidity taker:
\begin{equation*}
    (X_{-} + (1-\tau H(\Delta X)) \Delta X)^\alpha(Y_{-}+ (1-\tau H(\Delta Y))\Delta Y)^{1-\alpha} = X_-^\alpha Y_{-}^{1-\alpha}.
\end{equation*}
It is small in the sense that the no-arbitrage condition $B \leq S^\ast \leq A$ remains valid.
The second type is by a \emph{large order} from a liquidity taker and an optimal order from an arbitrageur that follows immediately. 
It is large in the sense that it creates an arbitrage opportunity, which is exploited immediately due to a competition among arbitrageurs.
We assume that these two transactions occur sequentially but at the same time, say, at time $t$, and regard $(\Delta X_t, \Delta Y_t)$ as the resulting change of the reserves due to the two transactions in opposite directions.
By Proposition~\ref{prop1}, we have
\begin{equation}\label{jump}
    (X_{-} + (1-\tau H(\Delta X)) \Delta X)^\alpha(Y_{-}+ (1-\tau H(\Delta Y))\Delta Y)^{1-\alpha} \geq X_-^\alpha Y_{-}^{1-\alpha}.
\end{equation}
The no-arbitrage condition $B \leq S^\ast \leq A$ remains valid (see Section~\ref{OA}).
The third type is by \emph{orders from arbitrageurs} invoked by the violation of 
$B \leq S^\ast \leq A$ due to the continuous move of $S^\ast$.
By competition among arbitrageurs, an arbitrarily small violation is exploited immediately. 
As a result, the reserves are updated continuously according to 
$\mathrm{d}X = \mathrm{d}X^\uparrow - \mathrm{d}X^\downarrow$ and 
$\mathrm{d}Y = \mathrm{d}Y^\uparrow- \mathrm{d}Y^\downarrow$ with
\begin{equation} \label{eq:amm1}
    \mathrm{d}Y^\downarrow_t = B_t \mathrm{d}X_t^\uparrow, \ \ 
    \mathrm{d}Y^\uparrow_t =  A_t \mathrm{d}X_t^\downarrow,
\end{equation}
where, $X^\uparrow$, $ X^\downarrow$,
$Y^\uparrow$ and $Y^\downarrow$ are continuous nondecreasing processes
satisfying
\begin{equation}\label{localtime}
X^\uparrow_t = \int_0^t 1_{\{B_u = S^\ast_u\}} \mathrm{d}X^\uparrow_u,\ \ 
X^\downarrow_t = \int_0^t 1_{\{A_u = S^\ast_u\}} \mathrm{d}X^\downarrow_u.
\end{equation}
The last equations describe that $X^\uparrow$ (resp. $X^\downarrow$) increases  only when $B = S^\ast$ (resp. $A = S^\ast$).
Consequently,  we have
\begin{equation}\label{dec}
\begin{split}
    &X_t = X^c_t + \sum_{s \in (0,t]} \Delta X_s, 
    \ \ X^c_t = X_0 + X^\uparrow_t - X^\downarrow_t,
    \\
     &Y_t = Y^c_t + \sum_{s \in (0,t]} \Delta Y_s, 
     \ \ Y^c_t = Y_0 + Y^\uparrow_t - Y^\downarrow_t
\end{split}
\end{equation}
with \eqref{jump}, \eqref{eq:amm1} and \eqref{localtime}.
\\

Note that $B \leq S^\ast \leq A$  is equivalent to
\begin{equation}\label{NA2}
    -\tau S \leq S^\ast - S \leq \frac{\tau}{1-\tau}S.
\end{equation}
Even if $S^\ast$ is a martingale of nondegenerate quadratic variation, $S$ is not. 
In fact, since $X$ and $Y$ are of finite total variation, $S$ enjoys this property as well. 
Between two consecutive jumps, $(X,Y)$ moves continuously with \eqref{NA2}.
Such a stochastic system is well-defined by the following theorem.
\begin{thm}\label{propSk}
Assume $S^\ast$ to be a positive continuous process. 
For any $x > 0$ and $y >0$ with
\begin{equation}\label{NA3}
     -\tau S_0 \leq S^\ast_0 - S_0 \leq \frac{\tau}{1-\tau}S_0
\end{equation}
for $S_0 = \beta y/x$,
there exist continuous processes of finite variation
 $X = x+ X^\uparrow - X^\downarrow$ with $X_0 = x$ and $Y = y + Y^\uparrow- Y^\downarrow$ with $Y_0=y$ such that \eqref{eq:amm1}, \eqref{localtime}
and \eqref{NA2} hold. 
\end{thm}
{\it Proof:} The result follows from the solvability of the Skorokhod problem; see Appendix~\ref{appB}.\\

\section{Impermanent Loss}\label{sec:imploss}
The value of the LP is defined by $V :=  Y + X S^\ast$.
\begin{thm}\label{thm2}
If $S^\ast$ is a continuous semimartingale, then
 \begin{equation*}
 V_T \geq V_0 + \int_0^T X_t\mathrm{d}S^\ast_t 
 \end{equation*}
 for any $T\geq 0$. The equality holds when $X = X_0 +X^\uparrow - X^\downarrow$ with \eqref{localtime}.
\end{thm}
{\it Proof:}
 By integration-by-parts, since $S^\ast$ is continuous and $X$ is of finite variation,
 \begin{equation}\label{IBP}
 V_T = V_0 + \int_0^T X_t\mathrm{d}S^\ast_t + \int_0^T \mathrm{d}Y^c_t
 + \int_0^T S^\ast_t \mathrm{d}X^c_t + 
\sum_{t \in (0,T]} (\Delta Y_t + S^\ast_t \Delta X_t ) 
 \end{equation}
 from \eqref{dec}.
 Using that $S^\ast = S^\ast_{-}$, we also have
 \begin{equation*}
 (1-\tau)\frac{\beta Y_-}{X_-} = B_- \leq S^\ast \leq A_- =\frac{1}{1-\tau} \frac{\beta Y_-}{X_-}.
 \end{equation*}
 Therefore from \eqref{jump},
 \begin{equation*}
     \begin{split}
         \Delta Y + S^\ast \Delta X &= 
         (\Delta Y)^+ - S^\ast (\Delta X)^- - (\Delta Y)^- + S^\ast (\Delta X)^+
         \\
         & \geq   \frac{Y_-}{1-\tau}\left( 
         (1-\tau)\frac{(\Delta Y)^+}{Y_-} - \beta \frac{(\Delta X)^-}{X_-}
         \right)
         +   Y_-\left( -\frac{(\Delta Y)^-}{ Y_-} + 
         \beta (1-\tau)\frac{(\Delta X)^+}{X_-}  
         \right) \\
         & \geq 
         \frac{Y_-}{1-\tau}\left( 
         \left(1 - \frac{(\Delta X)^-}{X_-}\right)^{-\beta} -1  - \beta \frac{(\Delta X)^-}{X_-}
         \right)\\
          & \hspace{1cm} Y_-\left( 
          \left(1 +(1-\tau) \frac{(\Delta X)^+}{X_-}\right)^{-\beta} -1  + 
         \beta (1-\tau)\frac{(\Delta X)^+}{X_-}  
         \right) 
         \\
         & \geq 0.
     \end{split}
 \end{equation*}
Here we have used that $z \mapsto (1+z)^{-\beta}-1 + \beta z$ is increasing for $z>0$ and is decreasing for $z < 0$.
We also have
\begin{equation*}
    \mathrm{d}Y^\uparrow_t - S^\ast_t \mathrm{d}X^\downarrow_t = 
     \mathrm{d}Y^\uparrow_t - S^\ast_t 1_{\{A_t = S^\ast_t\}}\mathrm{d}X^\downarrow_t =
     \mathrm{d}Y^\uparrow_t - A_t \mathrm{d}X^\downarrow_t = 0
\end{equation*}
and
\begin{equation*}
   - \mathrm{d}Y^\downarrow_t + S^\ast_t \mathrm{d}X^\uparrow_t = 
   -  \mathrm{d}Y^\downarrow_t + S^\ast_t 1_{\{B_t = S^\ast_t\}}\mathrm{d}X^\uparrow_t =
    - \mathrm{d}Y^\downarrow_t + B_t \mathrm{d}X^\uparrow_t = 0
\end{equation*}
from \eqref{eq:amm1} and \eqref{localtime}. \hfill{////}\\

\begin{cor}
    Under the assumption of Theorem~\ref{thm2}, the IL $\Psi_T := Y_0 + X_0S^\ast_T- V_T$ is super-hedged by a model-free rebalancing strategy in $S^\ast$ with zero cost:
    \begin{equation}\label{eq:psiApprox}
        \Psi_T  \leq \int_0^T (X_0-X_t)\mathrm{d}S^\ast_t.
    \end{equation}
\end{cor}

\begin{rem}\label{rem1}\upshape
The dynamics of IL under no fee ($\tau = 0$) are quite different. 
First, note that the bid-ask spread is degenerate in this situation, and thus $S^\ast_t = S_t$ in the absence of arbitrage.
Second, 
$L:= X^\alpha Y^{1-\alpha}$ is constant and hence,
\begin{equation*}
    Y_t = L \left(\frac{Y_t}{X_t}\right)^\alpha = 
    \frac{L}{\beta^\alpha} S_t^\alpha,
\end{equation*}
which in turn implies
\begin{equation*}
    \Psi_t = Y_0 + X_0 S_t - \frac{1+\beta}{\beta^\alpha} L S_t^\alpha = :\psi(S_t).
\end{equation*}
Note that $\psi$ is strictly convex, which implies a positive expected IL \eqref{Jensen}.
When $S^\ast$ is a positive continuous local martingale as in the Black-Scholes model, 
\begin{equation*}
\begin{split}
    \Psi_T &= \int_0^T \psi^\prime(S_t)\mathrm{d}S_t +
    \frac{1}{2}  \int_0^T \psi^{\prime\prime}(S_t)\mathrm{d}\langle S \rangle_t
    \\& =  \int_0^T (X_0-X_t)\mathrm{d}S^\ast_t +
    \frac{1}{2} (1-\alpha)^\alpha \alpha^{1-\alpha} L  \int_0^T (S^\ast_t)^\alpha \mathrm{d}\langle \log S^\ast \rangle_t
\\ & \geq \int_0^T (X_0-X_t)\mathrm{d}S^\ast_t
    \end{split}
\end{equation*}
by It\^o's formula.
The second term of the second line is the Loss-Versus-Rebalancing (LVR) in the terminology of~\cite{MMRZ}. 
Notice that the source of the LVR is the quadratic variation of $S = S^\ast$.
Under $\tau > 0$, the no-arbitrage constraint is merely $(1-\tau)S \leq S^\ast \leq S/(1-\tau)$, and indeed $S$ is of finite variation (hence, zero quadratic variation). 
This explains why there is no LVR under positive transaction fees, however small. 
\begin{cor}
    As a consequence of equation~\eqref{eq:psiApprox}, there is no LVR in the presence of a proportional transaction fee $\tau > 0$.
\end{cor}
\noindent

\end{rem}

\begin{rem}\upshape
    When $\tau > 0$ is small, $S^\ast \approx S$ and so
    $V = Y + XS^\ast \approx XS^\ast/\alpha$. Therefore,
    \begin{equation*}
        \int_0^T X_t \mathrm{d}S^\ast_t \approx 
        \int_0^T \alpha \frac{V_t}{S^\ast_t} \mathrm{d}S^\ast_t, 
    \end{equation*}
    meaning that $\alpha V$ units of the total wealth $V$ are invested in $S^\ast$.
    Our hedging strategy~\eqref{eq:psiApprox} is therefore to go long a buy-and-hold portfolio and short a constant-weights portfolio\footnote{Note, however, that $V$ is a self-financing wealth process only in the case where there is no liquidity taker, by Theorem~\ref{thm2}.}.
    Because such a constant-weights portfolio strategy is the optimal investment under a suitable framework\footnote{E.g., the Kelly portfolio under logarithmic utility~\cite{TK}}, it  also makes sense for the liquidity provider to leave the IL unhedged and regard the liquidity provision as an alternative way of asset management.
\end{rem}

\begin{rem}
\upshape
The liquidity provider herself can play the role of an arbitrageur. 
For example, if the external exchange rate equals the LP's bid price (i.e., 
$S^\ast_t = B_t$), an arbitrageur buys the first currency ($X$) from the external market at the price $S^\ast$ and swaps it for the second currency ($Y$) in the LP. 
On the other hand, the super-hedging strategy for the liquidity provider is to \emph{sell} the same amount as the increment of the first currency at the external market.
The activities of the arbitrageur and the liquidity provider therefore offset at the external market.
If the arbitrageur is the liquidity provider herself, the transactions at the external market can be bypassed, and the strategy amounts to reallocating the currency pairs between the LP and her hedging account. 
This approach avoids both discretization errors for continuous hedging and possible transaction costs in the external markets. 
In summary, a liquidity provider should act as an arbitrageur in order not to allow arbitrage opportunities to arise and be exploited by other arbitrageurs.
\end{rem}

\begin{rem}
    \upshape
    The incremental IL associated with
    the reserve update $(\Delta X_t,\Delta Y_t)$ due to a small liquidity taker's order is super-hedged by selling $\Delta X_t$ units of the first currency at the external market.
    This is actually a trivial arbitrage trade.
    Indeed, when $\Delta X_t > 0$, the exchange ratio is dominated as
    \begin{equation*}
        -\frac{\Delta Y_t}{\Delta X_t}
        < B_t \leq S_t^\ast,
    \end{equation*}
    while the super-hedging strategy sells the amount $\Delta X_t$ of the first currency for $S^\ast_t \Delta X_t$ units of the second currency at the external market. The arbitrage profit  for the liquidity provider is
    \begin{equation*}
       S^\ast_t \Delta X_t - (-\Delta Y_t) > 0 
    \end{equation*}
    in terms of the second currency.
\end{rem}

\section{Simulation}\label{sec:sim}
In this section, we perform simulations using historical price data to represent the external market price $S^*$.
We consider three different agents: small traders (the bid-ask spread after a small trade contains the external market price $S^*$), large traders (the resulting bid-ask spread creates an arbitrage opportunity), and arbitrage traders that perform the optimal arbitrage trade (Section~\ref{OA}) whenever $S^*$ lies outside the bid-ask range.

\begin{figure}[h]
\centering
\includegraphics[width=0.95\textwidth]{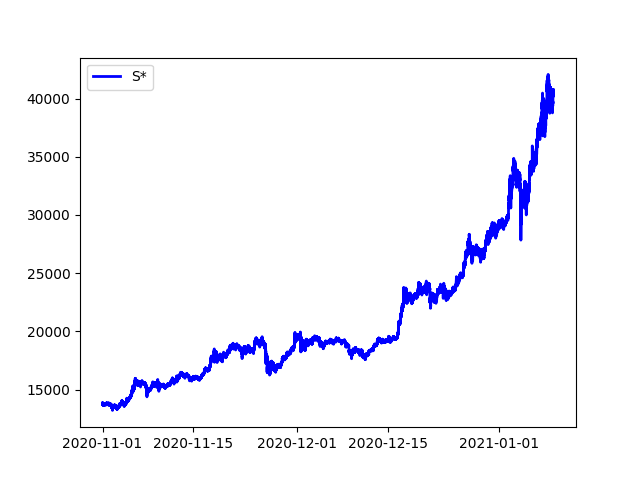}
\caption{\textbf{External  Price Data.}
Our external market price data consists of historical Bitcoin/USD mid-price data from 2020-11-01 to 2021-08-10 at 30-second intervals (200,000 observations).
We chose a period dominated by a price rally to obtain a potentially large
Impermanent Loss.\label{fig:px}}
\end{figure}

Our external market price data consists of historical Bitcoin/USD index-price data from BitMEX~\cite{BitMEX}, ranging from 2020-11-01 to 2021-08-10 at 30-second intervals 
(200,000 observations).\footnote{We obtained data for this study from BitMEX, a cryptocurrency trading platform. We gratefully acknowledge BitMEX for providing this data. BitMEX data is
publicly available at \href{http://public.bitmex.com}{public.bitmex.com}} 
Prices start at \$13,771 and end at \$45,601 with minimum at \$13,216.5 and maximum at \$42,076.5. Figure~\ref{fig:px}
plots the data. We assume that the liquidity pool starts with $\$1,000,000$ in USD and $72.616$ BTC such that the initial exchange rate equals the initial external price $S^*$ for $\alpha=0.5$.

\subsection{Discretization}
The external price $S^*$ is observed at discrete time steps $i\Delta t$: $S^*_{i\Delta t}$, where $\Delta t$ is a constant and $i\in \mathbb{N}^+$. 
The agents are small traders, large traders, and arbitrage traders, as defined in Section~\ref{sec:reserves}.

If the observation of $S^*$ falls outside the pool bid-ask spread, an arbitrage trader performs an optimal trade and thereby ensures that the observed $S^*$ falls within the bid-ask spread.
If no-arbitrage trade is required, a liquidity taker executes a trade with probability $p$. Given there
is a liquidity taker trade executed, there is a probability $p_S$ that it is a small trade.
Large trades are immediately followed by an optimal arbitrage trade such that the no-arbitrage condition holds. 

At each discrete time step, (i) trades occur as specified above, and (ii), we adjust the hedge using
the discretized version of the upper bound of Eq.~\eqref{eq:psiApprox} to the IL $\Psi_T$:
\begin{align*}
    \hat{\Psi}_{i} \gets \hat{\Psi}_{i-1} + 
    (x_0 - x_{(i-1)\Delta t})  \left(S^*_{i\Delta t} - S^*_{(i-1)\Delta t}\right),
\end{align*}
where $x_{(i-1)\Delta t}$ reflects the reserves before the trades of the current period, and $S^\ast_{(i-1)\Delta t}$ reflects the external
price at the beginning of the period.
The reserves $x_{(i-1)\Delta t}$ do not necessarily imply a bid-ask spread that fulfills the no-arbitrage condition at time $i\Delta t$. However, Equation~\eqref{eq:psiApprox} assumes that the no-arbitrage condition holds and therefore the inequality can be violated
in the discretized setting.
    \nothing{\footnote{In Appendix~\ref{appSim} we show the results under the assumption that the hedge adjustment follows after the no-arbitrage trade and before the next external price observation, in which case the super-hedge property is maintained. However, we consider this setup less realistic.}}

We provide a pseudo-code for the simulation algorithm, Algorithm~\ref{tab:algorithm} for different fees and different agents.

\subsection{Experiments}
We perform three different experiments with fees $\tau$ of 1, 5, 10, 15,  or 30 basis points, respectively.
The first experiment consists of arbitrage traders and small traders. 
The latter trade only if a no-arbitrage trade occurred in the same time slot.
In the second experiment, we replace the small trades by large trades that are immediately followed by optimal arbitrage trades (in the same discrete time step). 
The last experiment assumes there are only arbitrage
traders and the only movement in the reserve pool is due to arbitrage traders.
Table~\ref{tab:sim} presents the results.

The IL is generally lower with more trades (larger trading probability $p$), since the AMM collects more fees.
Therefore, we report the IL divided by the number of total trades in column "IL/\#T". Negative numbers denote gains.

We report the estimation error "RelErr" as the hedge-error relative to the buy-and-hold portfolio value
at the end of the period: $(\hat{\Psi}_T - {\Psi}_T)/(V_T + {\Psi}_T)$, in percentage points. 
Negative values mean that the super-hedge property no longer holds.

We first investigate the IL.
Comparing the first and second experiment, small traders ($p=1, p^{(S)}=1$) versus large traders  ($p=1, p^{(S)}=0$), we see that for a given fee the IL per trade (${\Psi}_T/\# T$) is smaller if we have large traders. 
We can explain the higher AMM profit with Proposition~\ref{prop1} and also because the trades are larger and thus more fees are collected. 
Note that in the second experiment, we have exactly one arbitrage trader per $\Delta S_t$ observed, hence the number 200,000 is equal to the number of time steps. 
This is because in one time step, we either have an arbitrage trade, or we have a large trade that is followed by an arbitrage trade.
Analogously, the number of arbitrage trades and small trades add up to 200,000 in each row of the first experiment, because we have either an arbitrage trade or a small trade.

The third experiment with only arbitrage traders ($p=0, p^{(S)}=0$) shows that the higher the fees, the fewer the total trades -- a direct consequence of the larger bid-ask spread implied by the larger fee. 
When dividing the IL by the number of trades, we observe an increasing number $\Psi_T/\#T$ as the fee increases. 
Uniswap, for example, set trading fees at 30 basis points, at the time of writing.\footnote{\url{ https://docs.uniswap.org/contracts/v2/concepts/advanced-topics/fees}, accessed on Jan 31, 2022.}

We now investigate the approximation, $\hat{\Psi}_T$ and "RelErr". 
The relative error is negative if the discrete super-hedge approximation $\hat{\Psi}_T$ loses its super-hedge property and under-hedges.

In the first two experiments, small traders ($p=1, p^{(S)}=1$) versus large traders ($p=1, p^{(S)}=0$), we observe that the hedge-error when under-hedging is at most 1\%. 
With larger fees, the hedge effectiveness increases and there is indeed a super-hedge in the presence of liquidity takers and more realistic fees. 
In the absence of liquidity takers,
the hedge is not a perfect super-hedge (negative "RelErr"), but reasonably close.

\section{Conclusion}
We have derived an upper bound for Impermanent Loss for Geometric Mean Markets with fees. 
This upper bound holds in continuous time. 
In discrete time, when the no-arbitrage condition cannot be guaranteed, this strategy can lead to becoming under-hedged.
However, the simulations confirm that under reasonable fees and with active liquidity takers, the hedging strategy is indeed a super-hedge.



\begin{algorithm}
\caption{Simulation}
\begin{algorithmic}[1]

\Procedure{Sim}{$S,\tau,x_0,y_0,p,p^{(S)},S\_$}
    \State $t \gets 0$
    \State $\hat{\Psi} \gets 0$
    \State $x\_ \gets x_0$
    \While{$t<T$}
            \If{$S[t] \notin [bid, ask]$}
                \State executeOptimalArbitrageTrade()
            \ElsIf{randUniform()$<p$}\Comment{small liquidity taker trade}
                \If{$S[t]-bid > ask-S[t]$}\Comment{trade in dir with more leeway}
                    \State $\Delta y \gets $ MaxSmallBuy()$-\epsilon$
                    \State Swap(.,$\Delta y$) \Comment{Swap updates reserves x, y}
                \Else
                    \State $\Delta x \gets $ MaxSmallSell()$-\epsilon$
                    \State Swap($\Delta x$,.)
                \EndIf
            \Else   \Comment{large liquidity taker trade}
                \If{randUniform()$<0.5$}
                    \State Swap(., 2 MaxSmallBuy())
                \Else
                    \State Swap(2 MaxSmallSell(), .)
                \EndIf
                \State executeOptimalArbitrageTrade()
            \EndIf
            \State $\hat{\Psi}=\hat{\Psi} + (x_0 - x\_ )  (S[t] - S\_)$
            \State $S\_ \gets S[t]$
            \State $x\_ \gets x$
            \State $t \gets t+1$
    \EndWhile
\EndProcedure

\end{algorithmic}\label{tab:algorithm}
\end{algorithm}

\begin{table}[h]
    \centering
    \begin{tabular}{rrrrrrrrrrr}
\toprule
 $\tau$ &  $p$ &  $p^{(S)}$ &      $V_T$ &          ${\Psi}_T$ &   $\hat{\Psi}_T$ &   \#Arb &  \#Large &  \#Small &  ${\Psi}_T/\#T$ &  RelErr \\
\midrule
     1 &    1 &          1 &  3.442 &       0.506 &           0.467 & 144,543 &       0 &  55,457 &    2.5 &   -1.0 \\
     5 &    1 &          1 &  3.496 &       0.453 &           0.450 & 102,398 &       0 &  97,602 &    2.3 &   -0.1 \\
    10 &    1 &          1 &  3.645 &       0.304 &           0.403 &  93,269 &       0 & 106,731 &    1.5 &    2.5 \\
    15 &    1 &          1 &  3.905 &       0.044 &           0.324 &  91,545 &       0 & 108,455 &    0.2 &    7.1 \\
    30 &    1 &          1 &  5.660 &      -1.711 &          -0.179 &  90,748 &       0 & 109,252 &   -8.6 &   38.8 \\
\hline
     1 &    1 &          0 &  3.443 &       0.506 &           0.467 & 200,000 &  54,335 &       0 &    2.0 &   -1.0 \\
     5 &    1 &          0 &  3.524 &       0.424 &           0.442 & 200,000 &  98,125 &       0 &    1.4 &    0.4 \\
    10 &    1 &          0 &  3.774 &       0.175 &           0.367 & 200,000 & 107,982 &       0 &    0.6 &    4.9 \\
    15 &    1 &          0 &  4.213 &      -0.264 &           0.239 & 200,000 & 109,834 &       0 &   -0.9 &   12.7 \\
    30 &    1 &          0 &  7.527 &      -3.578 &          -0.669 & 200,000 & 
    110,922 &       0 &  -11.5 &   73.7 \\
\hline
     1 &    0 &          0 &  3.441 &       0.507 &           0.467 & 142,985 &       0 &       0 &    3.5 &   -1.0 \\
     5 &    0 &          0 &  3.457 &       0.491 &           0.462 &  86,334 &       0 &       0 &    5.7 &   -0.7 \\
    10 &    0 &          0 &  3.467 &       0.482 &           0.459 &  56,091 &       0 &       0 &    8.6 &   -0.6 \\
    15 &    0 &          0 &  3.472 &       0.477 &           0.458 &  40,816 &       0 &       0 &   11.7 &   -0.5 \\
    30 &    0 &          0 &  3.480 &       0.468 &           0.455 &  21,708 &       0 &       0 &   21.6 &   -0.4 \\
\bottomrule

\end{tabular}
\centering
    \caption{\textbf{Simulation Results.} 
    The columns show the following numbers:
    $\tau$ is the fee in basis points (=1e-4), $p$ the probability for a trade to occur if there is a no-arbitrage trade due to external price movements within the discrete time-slot, $p^{(S)}$ the probability for that trade to be a small trade, $V_T$ the LP portfolio value (\$) at the end of the time horizon, ${\Psi}_T$ the Impermanent Loss ($1e6$\$) -- negative numbers are gains, $\hat{\Psi}_T$ the discrete estimate of the upper bound of the Impermanent Loss ($1e6$\$), \#Arb the number of arbitrage trades, followed by the number of large and small trades, ${\Psi}_T/\#T$ the Impermanent Loss (\$) divided by the total number of trades.
    We define RelErr (\%) as the estimation error relative to the buy-and-hold portfolio value: $(\hat{\Psi}_T - {\Psi}_T)/(V_T + {\Psi}_T)$.\\
    }
    \label{tab:sim}
\end{table}

\clearpage
\begin{appendix}
\section{Proof of Proposition~\ref{prop1}}\label{appA}
We have four cases: 1) $\Delta x_1 > 0$, $\Delta x_2 < 0$ and $\Delta x>0$,
2) $\Delta x_1 < 0$, $\Delta x_2 > 0$ and $\Delta x>0$,
3) $\Delta x_1 > 0$, $\Delta x_2 < 0$ and $\Delta x<0$,
and 4)  $\Delta x_1 < 0$, $\Delta x_2 > 0$ and $\Delta x<0$.\\
Case 1): By definition, we have
\begin{equation*}
    (x + (1-\tau)\Delta x_1)^\alpha (y+\Delta y_1)^{1-\alpha} = x^\alpha y^{1-\alpha}
\end{equation*}
and
\begin{equation*}
    (x + \Delta x_1 + \Delta x_2)^\alpha (y+\Delta y_1 + (1-\tau)\Delta y_2)^{1-\alpha} = (x+\Delta x_1)^\alpha (y+\Delta y_1)^{1-\alpha}.
\end{equation*}
Then, since $\Delta y_2 > 0$
\begin{equation*}
    \begin{split}
   & (x + (1-\tau)\Delta x)^\alpha (y+\Delta y_1 + \Delta y_2)^{1-\alpha} \\ & =
    \frac{(x + (1-\tau)\Delta x)^\alpha}{(x +\Delta x)^\alpha }
    (x +\Delta x)^\alpha  (y+\Delta y_1 + \Delta y_2)^{1-\alpha}
\\
& > \frac{(x + (1-\tau)\Delta x)^\alpha}{(x +\Delta x)^\alpha }
    (x +\Delta x)^\alpha  (y+\Delta y_1 + (1-\tau)\Delta y_2)^{1-\alpha}
    \\
    &=
    \frac{(x + (1-\tau)\Delta x)^\alpha}{(x +\Delta x)^\alpha } (x+\Delta x_1)^\alpha (y+\Delta y_1)^{1-\alpha}
    \\
    &=  \frac{(x + (1-\tau)\Delta x)^\alpha}{(x +\Delta x)^\alpha }
     \frac{(x +\Delta x_1)^\alpha }{(x + (1-\tau)\Delta x_1)^\alpha}
     (x + (1-\tau)\Delta x_1)^\alpha(y+\Delta y_1)^{1-\alpha}\\
     &=\frac{(x + (1-\tau)\Delta x)^\alpha}{(x +\Delta x)^\alpha }
     \frac{(x +\Delta x_1)^\alpha }{(x + (1-\tau)\Delta x_1)^\alpha}
     x^\alpha y^{1-\alpha}.
    \end{split}
\end{equation*}
Since $\Delta x < \Delta x_1$ and
\begin{equation*}
    z \mapsto \left(\frac{1 + (1-\tau)z}{1+z}\right)^\alpha
\end{equation*}
is decreasing, we have
\begin{equation*}
    \frac{(x + (1-\tau)\Delta x)^\alpha}{(x +\Delta x)^\alpha }
     \frac{(x +\Delta x_1)^\alpha }{(x + (1-\tau)\Delta x_1)^\alpha} > 1
\end{equation*}
and so,
\begin{equation*}
  (x + (1-\tau)\Delta x)^\alpha (y+\Delta y_1 + \Delta y_2)^{1-\alpha} 
  >  x^\alpha y^{1-\alpha} = (x + (1-\tau)\Delta x)^\alpha (y+\Delta y)^{1-\alpha}
\end{equation*}
from which the claimed inequality holds.\\
Case 2) By definition, we have
\begin{equation*}
    (x + \Delta x_1)^\alpha (y+(1-\tau)\Delta y_1)^{1-\alpha} = x^\alpha y^{1-\alpha}
\end{equation*}
and
\begin{equation*}
    (x + \Delta x_1 + (1-\tau) \Delta x_2)^\alpha (y+\Delta y_1 + \Delta y_2)^{1-\alpha} = (x+\Delta x_1)^\alpha (y+\Delta y_1)^{1-\alpha}.
\end{equation*}
Then, since $\Delta x_1 < 0$ and $\Delta y_1 > 0$,
\begin{equation*}
    \begin{split}
   & (x + (1-\tau)\Delta x)^\alpha (y+\Delta y_1 + \Delta y_2)^{1-\alpha} \\ & >
   (x +\Delta x_1 + (1-\tau)\Delta x_2)^\alpha (y+\Delta y_1 + \Delta y_2)^{1-\alpha}
   \\
   &= (x+\Delta x_1)^\alpha (y+\Delta y_1)^{1-\alpha} \\
   &> (x + \Delta x_1)^\alpha (y+(1-\tau)\Delta y_1)^{1-\alpha} \\ & = x^\alpha y^{1-\alpha} \\
   &= (x + (1-\tau)\Delta x)^\alpha (y+\Delta y)^{1-\alpha} 
   \end{split}
   \end{equation*}
from which the claimed inequality holds.\\
Case 3) We have $\Delta y_1 < 0$, $\Delta y_2 > 0$ and $\Delta y > 0$ this case.
By symmetry between $(x,\alpha)$ and $(y,1-\alpha)$, from the result for Case 2), we have
\begin{equation*}
  (y + (1-\tau)(\Delta y_1 + \Delta y_2))^{1-\alpha} (x+\Delta x_1 + \Delta x_2)^{\alpha} 
  >  x^\alpha y^{1-\alpha} = (y + (1-\tau)\Delta y)^{1-\alpha} (x+\Delta x)^{\alpha}
\end{equation*}
from which the claimed inequality holds.\\
Case 4) We have $\Delta y_1 > 0$, $\Delta y_2 < 0$ and $\Delta y > 0$ this case,
so as in Case 3), the result follows from Case 1) by symmetry.
\hfill{////}\\

\section{Proof of Theorem~\ref{propSk}}\label{appB}
Let $\psi = \log S^\ast - \log S_0- \log (1-\tau)$
and $a = -2 \log (1-\tau)$.
Then, $0 \leq \psi_0 \leq a$ by \eqref{NA3}.
Let
$(\phi,\eta)$ be the solution of the Skorokhod problem on $[0,a]$ for the continuous path $\psi$.
See \cite{KLRS} for an explicit solution.
By definition, 
we have a decomposition $\eta = \eta^\uparrow- \eta^\downarrow$
with nondecreasing continuous paths $\eta^\uparrow$ and $\eta^\downarrow$, 
\begin{equation*}
    \eta^\uparrow_t = \int_0^t  1_{\{\phi_s = 0\}}\mathrm{d}\eta^\uparrow_s, \ \ 
      \eta^\downarrow_t = \int_0^t  1_{\{\phi_s = a\}}\mathrm{d}\eta^\downarrow_s,
\end{equation*}
and
\begin{equation*}
0 \leq \phi = \psi + \eta \leq a.
\end{equation*}
Define $X^\uparrow$ and $X^\downarrow$ as the solution of the linear equation
\begin{equation*}
    \mathrm{d}X^\uparrow_t = \frac{1}{1+(1-\tau)\beta} (X^\uparrow_t- X^\downarrow_t) \mathrm{d}\eta^\uparrow_t,\ \ 
     \mathrm{d}X^\downarrow_t = 
   \left( 1 +   \frac{\beta}{1-\tau}\right)^{-1} (X^\uparrow_t- X^\downarrow_t)
     \mathrm{d}\eta^\downarrow_t
\end{equation*}
with $X^\uparrow_0 = 0$ and $X^\downarrow_0 = 0$.
Note that $X = x + X^\uparrow - X^\downarrow$  then solves
\begin{equation*}
    \mathrm{d}X_t = X_t 
    \left(\frac{1}{1+(1-\tau)\beta} \mathrm{d}\eta^\uparrow_t - 
    \left( 1 +   \frac{\beta}{1-\tau}\right)^{-1} 
     \mathrm{d}\eta^\downarrow_t\right)
\end{equation*}
with $X_0 = x > 0$ and so,
\begin{equation*}
     X_t = x\exp\left\{ 
   \frac{1}{1+(1-\tau)\beta} \eta^\uparrow_t - 
    \left( 1 +   \frac{\beta}{1-\tau}\right)^{-1} 
    \eta^\downarrow_t \right\}.
\end{equation*}
This ensures that $X$ is a positive continuous process.
Define $Y = y + Y^\uparrow - Y^\downarrow$ 
by \eqref{eq:amm1} with $Y_0=y$.
We have
\begin{equation*}
    \mathrm{d}Y_t = \mathrm{d}Y^\uparrow_t - 
    \mathrm{d}Y^\downarrow_t = -
    \beta Y_t\left( (1-\tau) \frac{\mathrm{d}X^\uparrow_t}{X_t} - \frac{1}{1-\tau} \frac{\mathrm{d}X^\downarrow_t}{X_t}\right),
\end{equation*}
which in turn implies
\begin{equation*}
   Y_t =  y\exp\left\{ \int_0^t (1-\tau)\beta \frac{\mathrm{d}X^\uparrow_s}{X_s} - \int_0^t \frac{1}{1-\tau} \beta \frac{\mathrm{d}X^\downarrow_s}{X_s}
\right\}.
\end{equation*}
This ensures that $Y$ is also a positive continuous process.
Now, we have
\begin{equation*}
    \begin{split}
        \mathrm{d} \phi_t & = \mathrm{d}\psi_t + \mathrm{d}\eta^\uparrow_t - \mathrm{d}\eta^\downarrow_t  \\
        &= \mathrm{d} \log S^\ast_t + 
        (1+(1-\tau)\beta)\frac{\mathrm{d}X^\uparrow_t}{X_t}-
        \left( 1 +   \frac{\beta}{1-\tau}\right)
        \frac{\mathrm{d}X^\downarrow_t}{X_t}
        \\
&= \mathrm{d} \log S^\ast_t + \frac{\mathrm{d}X_t}{X_t} - \frac{\mathrm{d}Y_t}{Y_t}\\
&= \mathrm{d}\left( \log \frac{S^\ast}{(1-\tau) S} \right)_t
\end{split}
\end{equation*}
Since $\phi_0 = \psi_0=
\log S^\ast_0 - \log S_0- \log (1-\tau)$, we then conclude
\begin{equation*}
\phi = \log \frac{S^\ast}{(1-\tau) S}.
\end{equation*}
Note that $0 \leq \phi \leq a$ is equivalent to \eqref{NA2}. \hfill{////}\\

\end{appendix}

\end{document}